\documentclass[runningheads]{llncs}
\usepackage{graphicx}
\usepackage{cite}
\begin{document}
\title{Comparing Community-aware Centrality Measures in Online Social Networks}
%
%
\author{Stephany Rajeh* \orcidID{0000-0002-7686-8506} \and
Marinette Savonnet \orcidID{0000-0003-0449-5277} \and
Eric Leclercq \orcidID{0000-0001-6382-2288} \and
Hocine Cherifi \orcidID{0000-0001-9124-4921}}
\authorrunning{S. Rajeh et al.}
\titlerunning{Comparing Community-aware Centrality Measures}
%
\institute{Laboratoire d’Informatique de Bourgogne - University of Burgundy, Dijon, France
\email{*stephany.rajeh@u-bourgogne.fr}}
\maketitle       
%
\begin{abstract}
Identifying key nodes is crucial for accelerating or impeding dynamic spreading in a network. Community-aware centrality measures tackle this problem by exploiting the community structure of a network. Although there is a growing trend to design new community-aware centrality measures, there is no systematic investigation of the proposed measures' effectiveness. This study performs an extensive comparative evaluation of prominent community-aware centrality measures using the Susceptible-Infected-Recovered (SIR) model on real-world online social networks. Overall, results show that K-shell with Community and Community-based Centrality measures are the most accurate in identifying influential nodes under a single-spreader problem. Additionally, the epidemic transmission rate doesn't significantly affect the behavior of the community-aware centrality measures.

\end{abstract}

\section{Introduction}

With the plethora of data flowing into online social networks, representing the main entities and their interactions is essential. Networks offer an ideal representation of such complex systems to investigate their structure and dynamics. Identifying influential nodes is crucial for many applications such as designing lucrative marketing campaigns, targeting terrorist attacks, controlling epidemic spreading, detecting financial risks, and extracting salient features from visual content \cite{lu2016vital,lasfar2000content, rital2002combinatorial, rital2005weighted, demirkesen2008comparison,hassouni2006hos, pastrana2006predicting}. Centrality is one of the main approaches employed to do so. Classically, centrality measures exploit the topology and dynamics of networks \cite{lu2016vital}. They can be classified into two main groups, namely local and global. The former uses the node's neighborhood, while the latter incorporates all of the network's information to quantify a node's influence. They can also be combined \cite{sciarra2018change,ibnoulouafi2018m}.

Many real-world networks contain densely connected zones that are loosely linked to each other. This so-called community structure is a ubiquitous feature in natural and artificial systems \cite{girvan2002community}. The network's structure and dynamics are significantly affected by communities \cite{nematzadeh2014optimal}. Recently developed centrality measures exploit this information to identify influential nodes \cite{ghalmane2019immunization, guimera2005functional, tulu2018identifying, gupta2016centrality, modvitality, zhao2015community, luo2016identifying}. We refer to them as ``community-aware" centrality measures. Unlike classical centrality measures, community-aware centrality measures differentiate between the node's intra-community links (links between nodes in the same community) and inter-community links (links between nodes in different communities). Intra-community links exert influence at the community level, while inter-community links exert influence at the network level \cite{rajeh2021characterizing}. The difference between community-aware measures is mainly based on how intra-community links and inter-community links are associated together. For example, Comm centrality \cite{gupta2016centrality} preferentially selects bridges over hubs by prioritizing inter-community links over intra-community links. Community-based Mediator \cite{tulu2018identifying} favors nodes with unbalanced intra-community and inter-community links.

With limited resources, it is essential to identify top influential nodes either for maximizing or for minimizing the diffusion in online social networks. The Susceptible-Infected-Recovered (SIR) model is commonly used to model disease and rumor spreading \cite{anderson1979population}. Starting with a small set of initial spreaders defined by a specific centrality measure, the goal is to evaluate its ability to reach the maximum outbreak size.

The SIR model has been widely used to investigate the behavior of various classical centrality measures \cite{kitsak2010identification, bucur2020top, liu2015core}. Studies on community-aware centrality measures examine either a small number of the proposed solutions in the literature or experiments are performed on a small sample of networks \cite{tulu2018identifying, zhao2015community, modvitality, gupta2016centrality, luo2016identifying}. Therefore, there is no consensus about the effectiveness of the most popular measures on online social networks, where communities are naturally prevalent \cite{traud2011comparing,  labatut2014identifying, gaisbauer2021ideological}. This paper aims to fill this gap. An extensive investigation of seven community-aware centrality measures is performed on ten real-world online social networks using the SIR diffusion model under a single-spreader scheme.

The paper is organized as follows. Section \ref{sec:CommAwareCent} introduces the community-aware centrality measures. Section \ref{sec:DataToolsAndMethods} presents the networks, the tools, and the methodology applied. Experimental results are provided in section \ref{sec:ExpResults}. The main findings are discussed in section \ref{sec:Discussion}. Finally, in section \ref{sec:Conc}, the conclusion is given.


\section{Community-aware Centrality Measures}
\label{sec:CommAwareCent}
In this section, we briefly recall the definitions of the seven community-aware centrality measures under test. Let $G(V, E)$ be an undirected and unweighted graph where $V$ is the set of nodes, $E$ is the set of edges, and $N=|V|$ is the size of the network. It is partitioned into $N_c$ non-overlapping communities where $c_k$ is $k$-th community.  A node $i$ possess $k_i^{intra}$ intra-community links and $k_i^{inter}$ inter-community links such that $k_i^{tot} = k_i^{intra} + k_i^{inter}$ represents its degree.  Note that if the community structure is unknown, a community detection algorithm is needed to uncover it.

\textbf{1. Community Hub-Bridge} \cite{ghalmane2019immunization} weights the intra-community links of a node by its community size. The inter-community links are weighted by the number of communities reached by the node. It is defined as follows:

\begin{equation}
\alpha_{CHB}(i) = |c_k| \times k_i^{intra} + NNC_i \times k_i^{inter}
\end{equation}

where $|c_k|$ is the size of the community of node $i$ and $NNC_i$ is the number of communities linked to node $i$.

\textbf{2. Participation Coefficient} \cite{guimera2005functional} gives more importance on the heterogeneity of the inter-community links of a node. If the node's links are uniformly distributed across the communities, its centrality value is one. It is defined as follows:
\begin{equation}
\alpha_{PC}(i) = 1 - \sum_{c=1}^{N_c} 
\left(
\frac{k_{i,c}}{k_i^{tot}}
\right)^2
\end{equation}

where $k_{i,c}$ is the number of links node $i$ has in a given community $c$.

\textbf{3. Community‑based Mediator} \cite{tulu2018identifying} uses entropy to quantify the node's importance through its intra-community and inter-community links. It is defined as follows:
\begin{equation}
\alpha_{CBM}(i) = H_i \times \frac{k_i^{tot}}{\sum_{i=1}^{N} k^{tot}_i}
\end{equation}

where $H_i$=[$-\sum \rho_i^{intra} log(\rho_i^{intra})]$+$[- \sum \rho_i^{inter} log(\rho_i^{inter})$] is the entropy of node $i$ based on its $\rho^{intra}$ and $\rho^{inter}$ which represent the node's ratio of intra-community and inter-community links.

\textbf{4. Comm Centrality} \cite{gupta2016centrality} weights the intra-community links and inter-community links by the ratio of external links. It also prioritizes bridges over hubs. It is defined as follows:
\begin{equation}
\alpha_{Comm}(i) =  (1 + \mu_{c_k}) \times \chi +  (1 - \mu_{c_k})  \times  \varphi^2
\end{equation}

where $\mu_{c_k}$ is the proportion of inter-community links over the total community links in community $c_k$, $\chi = \frac{k_i^{intra}}{max_{(j \in c)}k_j^{intra}} \times R$, $\varphi = \frac{k_i^{inter}}{max_{(j \in c)}k_j^{inter}} \times R$, and $R$ is a constant to scale intra-community and inter-community values to the same range.

\textbf{5. Modularity Vitality} \cite{modvitality} is based on the modularity variation due to the node removal from the network. Removal of a bridge node increases the modularity, while removal of internal a hub decreases the modularity.  It is defined as follows: 
\begin{equation}
\alpha_{MV}(i) =   M(G) - M(G_i)
\end{equation}

where $M(G)$ is the modularity of a network and $M(G_i)$ is the network's modularity after the removal of node $i$. Note that Modularity Vitality is a signed centrality. In this study, we use its absolute value to rank the nodes.

\textbf{6. Community-based Centrality} \cite{zhao2015community} is based on weighting the node's intra-community and inter-community links by the subsequent sizes of their belonging communities. It is defined as follows:

\begin{equation}
\alpha_{CBC}(i) =  \sum_{c=1}^{N_c} k_{i,c} 
\left(
\frac{n_c}{N}
\right)
\end{equation}

where $n_c$ is the number of nodes in community $c$ and $k_{i,c}$ is the number of links node $i$ has in a given community $c$.

\textbf{7. K-shell with Community} \cite{luo2016identifying} is based on the $k$-shell (also called $k$-core) hierarchical decomposition of the network composed of intra-community links and the network composed of inter-community links, separately. A weighting parameter then combines the two values to prioritize the selection of hubs or bridges. It is defined as follows:

\begin{equation}
\alpha_{ks}(i) = \delta \times \alpha^{intra}(i) + (1- \delta) \times \alpha^{inter}(i) 
\end{equation}

where $\alpha^{intra}(i)$ and $\alpha^{inter}(i)$ stand for the $k$-shell value of node $i$ by only considering intra-community links and inter-community links, respectively. $\delta$ is set to 0.5 in this study.


\section{Data, Tools, and Methods}
\label{sec:DataToolsAndMethods}
\subsection{Data}
This study uses ten unweighted and undirected online social networks publicly available. They originate from various online platforms (Facebook, Twitter, Deezer, Hamsterster, and Pretty Good Privacy). Table \ref{TableBasicTopology} reports their basic topological characteristics. As their community structure is unknown, it is uncovered by Infomap \cite{rosvall2008maps}.

\textbf{1. Facebook Friends} \cite{netz}: Nodes are users from a Facebook ego network extracted in April 2014. Edges between two users mean they are ``friends" on Facebook.

\textbf{2. Retweets Copenhagen} \cite{nr}: Nodes are Twitter users tweeting while the United Nations conference in Copenhagen about climate change was taking place. Edges represent retweets.

\textbf{3. Caltech} \cite{nr}: Nodes are users on Facebook enrolled at Caltech University.  Edges between two users mean they are ``friends" on Facebook.

\textbf{4. Ego Facebook} \cite{nr}: Nodes are users on Facebook participating in a survey conducted on Facebook. Edges between two users mean they are "friends" on Facebook.

\textbf{5. Hamsterster} \cite{kunegis2014handbook}: Nodes represent users from an online social pet network hamsterster.com. Edges represent friendships between the users.

\textbf{6. Facebook Organizations} \cite{netz}: Nodes are users on Facebook who work in the same company. Edges between two users mean they are ``friends" on Facebook.

\textbf{7. Facebook Politician Pages} \cite{nr}: Nodes are Facebook pages of politicians from different countries. Edges represent mutual likes of Facebook users among the given pages.

\textbf{8. Princeton} \cite{nr}: Nodes are users on Facebook enrolled at Princeton University. Edges between two users mean they are ``friends" on Facebook.

\textbf{9. PGP} \cite{kunegis2014handbook}: Nodes are users from the web of trust, utilizing Pretty Good Privacy (PGP) encryption for sharing information online. Edges between users represent sharing data under secure connections.

\textbf{10. DeezerEU} \cite{rozemberczki2020characteristic}: Nodes represent users from Deezer, a European platform for music streaming. Edges represent online friendships between users.

\begin{table}[t]
\caption{Topological features of the networks. $N$ is the number of nodes. $|E|$ is the number of edges. $<k>$ is the average degree. $\zeta$ is the transitivity. $\mu$ is the mixing parameter. $\lambda{th}$ is the epidemic threshold. * means the largest connected component of the network is taken if it is disconnected.}
\begin{center}
\begin{tabular}{|p{2.5cm}|c|c|c|c|c|c|}
\hline
\textbf{Network} & $N$ & $|E|$ & $<k>$ & $\zeta$ & $\mu$ & $\lambda{th}$ \\
\hline

Facebook Fri.* & 329 & 1,954 & 11.88 & 0.512 & 0.112 & 0.048\\
Retweets Co. & 761 & 1,029 & 2.70 & 0.060 & 0.287 & 0.139 \\
Caltech* & 762 & 16,651 & 43.70 & 0.291 & 0.410 & 0.048 \\
Ego Facebook & 4,039 & 88,234 & 43.69 & 0.519 & 0.077 & 0.009 \\ 
Hamsterster* & 1,788  & 12,476 & 13.49 & 0.090 & 0.298 & 0.022\\
Facebook Org. & 5,524 & 94,219 & 34.11 & 0.222 & 0.366 & 0.016 \\
Facebook Pol. & 5,908 & 41,729 & 14.12 & 0.301 & 0.111 & 0.024 \\ 
Princeton* & 6,575 & 293,307 & 89.21 & 0.163 & 0.365 & 0.006\\
PGP & 10,680   & 24,316 & 4.55 & 0.378 & 0.172 & 0.056 \\
DeezerEU & 28,281  & 92,752 &  6.55 & 0.095 & 0.429 & 0.066 \\

\hline

\end{tabular}
\label{TableBasicTopology}
\end{center}
\end{table}

\subsection{Susceptible-Infected-Removed Model}
\label{sec:SIR}
The Susceptible-Infected-Removed (SIR) model is one of the widely used diffusion models in networks. Initially, a single node or a set of nodes ($f_o$) is in the infectious state (I) while the remaining nodes are in the susceptible state (S). At each iteration, an infectious node infects its susceptible neighbors at a rate $\lambda$. Previously infected nodes recover and are removed from the network at a rate $\gamma$. The spreading continues until there are no infectious nodes. At this point, the number of nodes in the ``Recovered" state indicates the spreading power of the single node or the initial set of nodes ($f_o$). Each network has an epidemic threshold ($\lambda_{th}$) controlling the epidemic spreading. It is defined as \cite{wang2016predicting}:

\begin{equation}
\lambda_{th} = \frac{<k>}{<k^2> -  <k>}
\end{equation}
where $<k>$ and $<k^2>$ are the first and second moments of the network's degree distribution. The epidemic threshold values are reported in table \ref{TableBasicTopology}.

\subsection{Imprecision function}
The imprecision function \cite{kitsak2010identification} measures the performance of a centrality measure in predicting influential spreaders. It is based on the average number of infections due to an infected seed node. It is defined as follows:

\begin{equation}
\epsilon_c(p) = 1 - \frac{M_c (p)}{M_{eff} (p)}
\end{equation}

where $p$ is a value between [0,1], $M_c (p)$ is the average spreading power of top $pN$ nodes ranked according to a specific centrality measure $c$, and $M_{eff} (p)$ is the average spreading power of top $pN$ nodes ranked according to their influence in the SIR model ($N$ is the number of nodes). The smaller the value of $\epsilon_c(p)$, the better the performance of the centrality measure $c$.

\subsection{Methods}
The SIR model runs on each network using different transmission rates around the epidemic threshold ($\frac{\lambda_{th}}{2}$, $\frac{\lambda_{th}}{1.5}$, $1.5 \times \lambda_{th}$, $2 \times \lambda_{th}$). The recovery rate $\gamma$ is set to 1 to measure the spreading ability of the seed node initiating the spreading only. For each transmission rate, 1000 independent simulations of the SIR model are performed in networks with less than 6000 nodes and 100 independent simulations otherwise. The SIR spread size of each node in the network is computed after setting it as the seed of diffusion. The set ordered from highest to smallest SIR spread size is called the reference set. The community-aware centrality measures are computed, and nodes are ranked from higher to lower centrality value. For each transmission rate ($\lambda$), we calculate the imprecision function over the top fraction $pN$ nodes.


\section{Experimental Results}
\label{sec:ExpResults}
\subsection{Performance of the community-aware centrality measures within networks}
\begin{figure*}[ht!]
\centerline{\includegraphics[width=5.5 in, height=3 in]{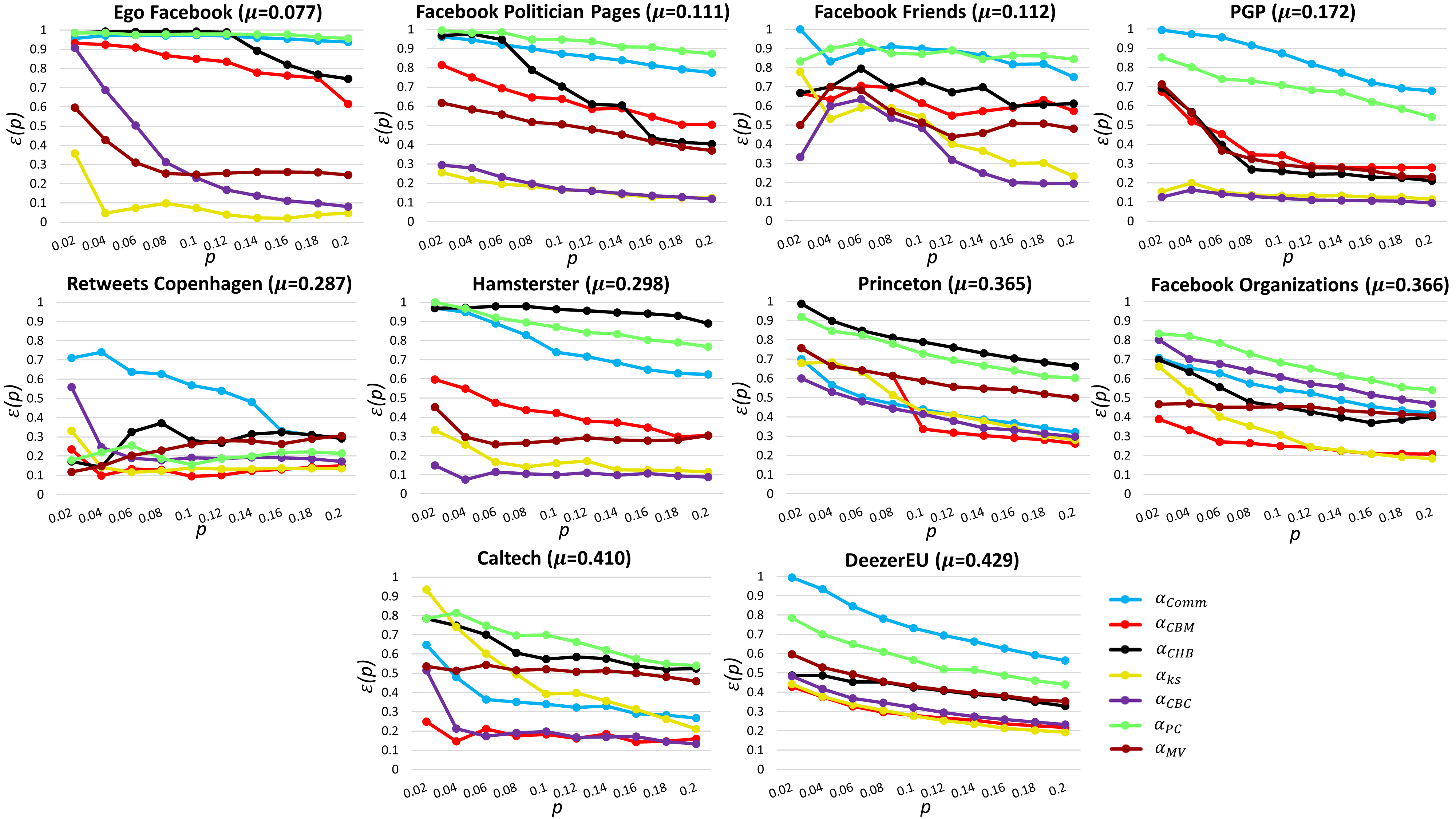}}
\caption{The imprecision function $\epsilon(p)$ for the 7 community-aware centrality measures on each network. The transmission rate is set to $\lambda_{th}$ and the recovery rate is set to 1. $\mu$ is the mixing parameter, the ratio of inter-community over total community links in a network. The community-aware centrality measures are: Comm Centrality = $\alpha_{Comm}$, Community‑based Mediator = $\alpha_{CBM}$, Community Hub-Bridge = $\alpha_{CHB}$, K-shell with Community = $\alpha_{ks}$, Community-based Centrality = $\alpha_{CBC}$, Participation Coefficient = $\alpha_{PC}$, and Modularity Vitality = $\alpha_{MV}$.}
\label{MainFigures}
\end{figure*}

Figure \ref{MainFigures} illustrates the performance of the community-aware centrality measures for the ten networks under study. The transmission rate is set equal to the epidemic threshold ($\lambda_{th}$) for each network. Each figure reports the evolution of the seven community-aware centrality measures' imprecision function when the top spreading nodes' size ranges from $p$=0.02 to $p$=0.2 of the network size. Remember that the lower the value of the imprecision function, the more effective the centrality measure. One can observe that the performance generally increases with the proportion of top spreading nodes. Furthermore, no community-aware centrality measure outperforms the others in all the situations.  Overall, there is a high variability of community-aware centrality measures performances within and across networks. For example, In Ego Facebook, at $p$=0.02, the imprecision value of K-shell with Community is 0.38, followed by Modularity Vitality at 0.6. Then all others have an imprecision value between 0.9 and 1. The variability among the community-aware centralities persists till $p$=0.2. K-shell with Community now has a value of 0.05, indicating its high accuracy at higher $p$. In the same vein comes Community-based Centrality, which has an imprecision value of 0.1. Its accuracy improves by almost 90\% compared to its value at $p$=0.02. Modularity Vitality follows with an imprecision value of 0.25, improving in almost half value of $\epsilon(p)$. Community-based Mediator improves from 0.92 ($p$=0.02) to a value of 0.61 ($p$=0.20). Community Hub-Bridge also improves, but in a lower proportion. Finally, Participation Coefficient and Comm Centrality show a negligible improvement. There is also a high variability for the same community-aware centrality measures across networks. For example, in Facebook Politician Pages, the imprecision value of Community-based Mediator at $p$=0.02 is 0.81, while in Caltech, it amounts to 0.25. Another example is Community-based Centrality in Ego Facebook amounting to 0.9 at $p$=0.02 while it amounts to 0.11 in PGP.

\begin{figure*}[ht!]
\centerline{\includegraphics[width=5.5 in, height=2 in]{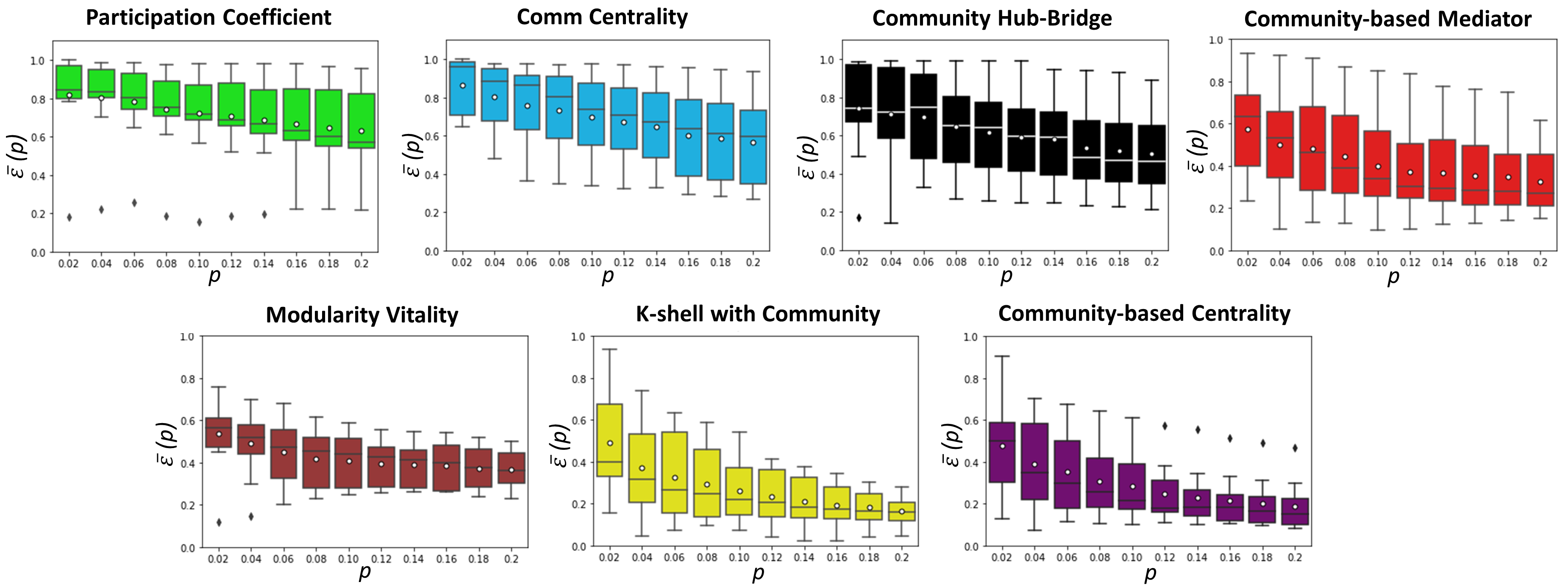}}
\caption{The average of the imprecision function $\overline{\epsilon}(p)$ over the 10 online social networks. The transmission rate is set to $\lambda_{th}$ and the recovery rate is set to 1.}
\label{BoxPlotofMainFigures}
\end{figure*}

\subsection{Performance of the community-aware centrality measures across networks}
Each community-aware centrality measure's imprecision function is averaged over the ten networks for all $p$ values. The goal is to better understand the performance consistency. Figure \ref{BoxPlotofMainFigures} illustrates these results. The most stable (low variability) community-aware centrality measure is Modularity Vitality. Despite the change in $p$, the imprecision function values remain stable and condensed. Then comes K-shell with Community and Community-based Centrality. Even though they show high variability when $p \leq 0.08$, both are very consistent afterward.
On the opposite, the remaining community-aware centrality measures show higher variability as $p$ increases. The average imprecision function $\overline{\epsilon}(p)$ illustrates the high accuracy of K-shell with Community and Community-based Centrality for all $p$ values. It ranges from 0.5 for the lowest $p$ value to 0.1 at the highest $p$ value. Then comes Modularity Vitality, with $\overline{\epsilon}(p)$ = 0.55 at $p$=0.02 and $\overline{\epsilon}(p)$ = 0.40 at $p$=0.20. Community-based Mediator has similar $\overline{\epsilon}(p)$ values as Modularity Vitality, yet it has high variability. Community Hub-Bridge shows $\overline{\epsilon}(p)$ between 0.75 and 0.5 at $p$=0.02 and $p$=0.20, respectively. Participation Coefficient and Comm Centrality perform poorly. Their minimum for $\overline{\epsilon}(p)$ is around 0.6, and their maximum is around 0.8. These results confirm the results of figure \ref{MainFigures}.

\subsection{Influence of the transmission rate}
\begin{figure*}[ht!]
\centerline{\includegraphics[width=5.5 in, height=3 in]{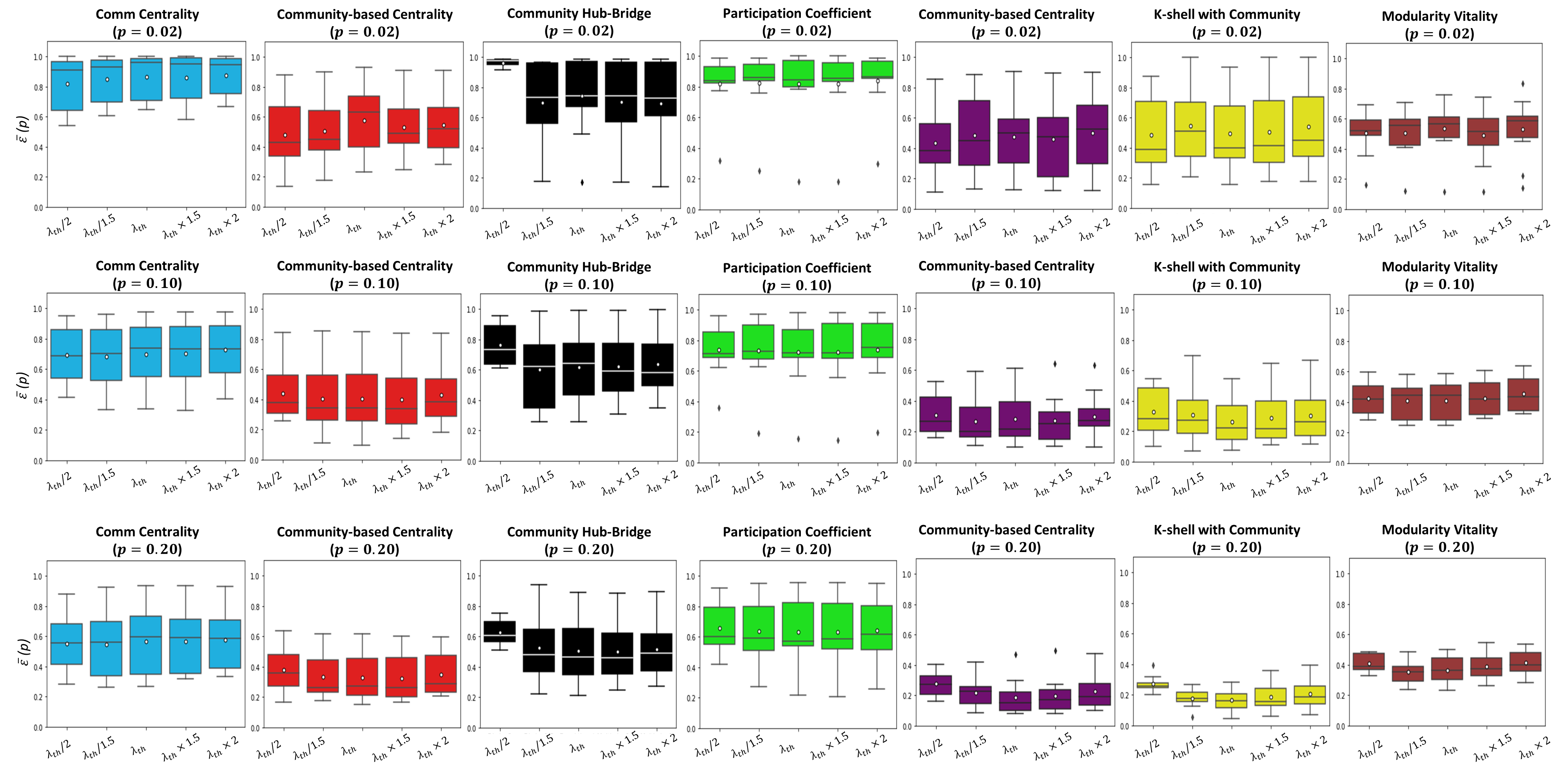}}
\caption{The average of the imprecision function $\overline{\epsilon}(p)$ over the 10 online social networks as a function of five different transmission rates ($\frac{\lambda_{th}}{2}$, $\frac{\lambda_{th}}{1.5}$, $\lambda_{th}$, $1.5 \times \lambda_{th}$, $2 \times \lambda_{th}$). The recovery rate is set to 1. The upper, middle, and bottom figures show the results at $p$=0.02, $p$=0.10, and $p$=0.20, respectively.}
\label{DifferentTransmissionRatesAnalysis}
\end{figure*}

In this experiment, we study the effect of varying the transmission rate ($\lambda$) in the SIR model around the epidemic threshold ($\lambda_{th}$). Figure \ref{DifferentTransmissionRatesAnalysis} shows the average imprecision function $\overline{\epsilon}(p)$ of the seven community-aware centrality measures at five different transmission rates. The average imprecision function $\overline{\epsilon}(p)$ is calculated considering a low portion of top nodes ($p$=0.02), a medium portion of top nodes ($p$=0.10), and a high portion of top nodes ($p$=0.20).

At low $p$ values ($p$=0.02), results are generally comparable. For example, the mean of $\overline{\epsilon}(p)$ for Comm Centrality at the five different transmission rates is in the vicinity of 0.8. Also, the boxplots' interquartile range is quite similar, indicating that the values are consistent across $\lambda$. Participation Coefficient, Community-based centrality, K-shell with Community, and Modularity Vitality also show consistent results. In contrast, Community Hub-Bridge is the most sensitive to the variation of the transmission rate. When $\lambda = \frac{\lambda_{th}}{2}$, Community Hub-Bridge cannot differentiate the nodes. Indeed, the mean $\overline{\epsilon}(p)$ is 0.98, and the interquartile range's height is very narrow. When the transmission rate is set to $\frac{\lambda_{th}}{1.5}$, $\lambda_{th}$, $1.5 \times \lambda_{th}$, and $2 \times \lambda_{th}$, $\overline{\epsilon}(p)$ becomes quite comparable.
The consistency of the mean and the interquartile range of  $\overline{\epsilon}(p)$ persist at $p$=0.10 and at $p$=0.20. Indeed, they share approximately the same values of $\overline{\epsilon}(p)$ for all community-aware centrality measures except for Community Hub-Bridge. Although now its interquartile range is wider compared to that of $p$=0.02 when $\lambda = \frac{\lambda_{th}}{2}$, the mean and interquartile range are much different than the others.

\section{Discussion}
\label{sec:Discussion}
This study aims to investigate the behavior of popular community-aware centrality measures in online social networks. Community-aware centrality measures quantify a node's importance based on its local influence (inside its community using intra-community links) and its global impact (outside of its community using inter-community links). Yet, each community-aware centrality measure processes these two types of links distinctively.

A comparative evaluation of seven community-aware centrality measures is performed using 
the SIR diffusion model under a single-spreader scheme. The imprecision function quantifies the centrality measure's accuracy by comparing the spreading power of top nodes according to a centrality measure compared to their ground truth spreading efficiency.  Results indicate that K-shell with Community and Community-based Centrality outperform the alternative community-aware centrality measures. K-shell with Community exploits the hierarchical structure of the networks while taking into consideration its community structure. This result corroborates the study reported in \cite{kitsak2010identification}.

Indeed, under a single-spreader setting, nodes identified using $k$-shell are the most accurate in predicting spreading outbreaks in networks. The performance of Community-based Centrality is also on the same line as the findings of the authors who proposed this measure \cite{zhao2015community}. This study shows that this community-aware centrality measure is accurate in online social networks with communities of different sizes.
Results also show that Community-based Mediator is somewhat sensitive to the community structure strength. Indeed, as shown in figure \ref{MainFigures}, when the network has a strong community structure ($\mu \leq 0.172$), it performs poorly. Yet, as the network has a weaker community structure, it becomes as accurate as K-shell with Community and Community-based Centrality. This centrality exploits the heterogeneity of links to assess the node's importance. Indeed, in a weak community structure, a node possesses a higher number of inter-community links than intra-community links.
It explains why it performs better in a weak community structure. Modularity Vitality is the most consistent across networks, regardless of the strength of the community structure. The low accuracy of Participation Coefficient, Comm Centrality, and Community Hub-Bridge may be linked to the fact that they give a lot of importance to bridges. Besides bridges, online social networks also include hubs inside their communities that play a major role in information dissemination.

\section{Conclusion}
\label{sec:Conc}

Identifying influential nodes in online social networks is fundamental for maximizing information diffusion and inhibiting fake news spreading. The community structure of a network plays a crucial role in the dynamics of these spreading processes. This work investigates the effectiveness of prominent community-aware centrality measures to target influential nodes using the SIR diffusion process under a single-spreader scheme. Results show that K-shell with Community and Community-based Centrality are the most accurate community-aware centrality measures. Additionally, performances are pretty insensitive to variation of the transmission rate. Therefore, this work gives clear indications about which community-aware centrality measure to use when resources are restrained to target single nodes. Nevertheless, practitioners need to be conscious that the community aware-centrality measures accuracy depends on the seed node size. As in numerous situations, the community structure is unknown. Future work will investigate the results consistency using alternative community detection algorithms. Another direction of research is to study the influence of the community-aware centrality measures using different propagation processes. Finally, we are planning to link the performances to the network topological properties.

\bibliographystyle{splncs04}
\bibliography{biblio}

\end{document}